# Hot interiors of ice giant planets inferred from electrical conductivity of dense H$_2$O fluid

Kenta Oka[1], Yoshiyuki Okuda[1]* & Kei Hirose[1,2]

[1]Department of Earth and Planetary Science, The University of Tokyo, Tokyo, Japan

[2]Earth-Life Science Institute, Tokyo Institute of Technology, Tokyo, Japan

**Uranus and Neptune have intrinsic magnetic fields generated via convection in a molten H$_2$O layer, where the field strength is determined by its electrical conductivity (EC) along with convection size and velocity[1]. Previous shock experiments[2–7] reported that the EC of molten H$_2$O is high enough to generate magnetic fields of these ice giant planets with adiabatic thermal structures[8]. Here we measured the EC of ionic H$_2$O fluid for the first time by static compression experiments up to 45 GPa and 2,750 K. The EC determined is lower by a few orders of magnitude than earlier data by shock compression measurements and not capable of generating a magnetic field with the conventional interior thermal structures. Our results necessitate recently-suggested fewfold hotter interiors[9–11] of Uranus and Neptune to explain their magnetic fields.**

The outermost solar system planets of Uranus and Neptune, referred to as ice giants, are thought to be composed mainly of H$_2$O, one of the most abundant molecules in the universe. Recent studies have revealed the phase diagram of H$_2$O ice[12,13] (Fig. 1), indicating that Uranus' and Neptune's mantle may consist of dense ionic H$_2$O fluid and underlying superionic H$_2$O ice XVIII when considering conventional, adiabatic thermal structures[8] without thermal boundary layers in the ice giants' interiors. While their magnetic fields are known to be as intense as that of the Earth, they show strong multipole components unlike the predominantly dipole one for our planet. Since the multipole components diminish with increasing distance more rapidly than the dipole ones, it has been believed that the magnetic fields of the ice giants are primarily generated near the planet's surface, most likely in the ionic H$_2$O fluid layer[14] (Fig. 1).

In order to generate and maintain the magnetic field by a dynamo action, the magnetic Reynolds number $R_m$ should be larger than fourty[1];

$$R_m = \mu_0 \sigma U L \qquad (1)$$

where $\mu_0$ is the permeability of free space, $\sigma$ is EC, and $U$ and $L$ are typical convection velocity and length scale, respectively. Previous EC measurements of ionic $H_2O$ fluid have been limited to shock compression studies with measurement time scales of $10^{-9}$–$10^{-6}$ seconds[2–7], all demonstrating the EC of about $10^3$ to $10^4$ S/m along the presumed isentropes[8] for the ice giants. On the other hand, recent models[9,10] argued that an adiabatic interior structure of Uranus without thermal boundary layers fails to explain its "faint" feature, requiring a few times hotter interior (Fig. 1). If this is the case, the total EC of $H_2O$ fluids in the ice giants is controlled by electronic conduction rather than ionic conduction. The opaque and electronically conducting liquid $H_2O$ has been observed based on its reflectivity at >~100 GPa and >~5,000 K during dynamical compression[15–17] or above 4,300 K at 17 GPa from recent absorption coefficient measurements under static compression in a diamond-anvil cell (DAC)[12].

Here we determine the EC of dense ionic $H_2O$ fluid by static experiments in the DAC at high pressure and temperature (*P-T*) (Fig. 1), which is important for understanding the origin of the magnetic fields of ice giants. While existing models[8–11] predicted temperatures for the molten $H_2O$ layer that vary by thousands of degrees (Fig. 1), the present results infer the thermal structures of the ice giants, which give clues for their formation processes and interior dynamics.

**High-pressure and -temperature electrical conductivity measurements**

We measured the EC of dense ionic $H_2O$ fluid under high *P-T* up to 45 GPa and 2,750 K via static compression experiments in a laser-heated DAC. The EC of $H_2O$ ice (solid) was also obtained at high *P-T* to be compared with that of the fluid. We employed a method recently developed[18] to measure the EC of a material that is transparent to a near-infrared laser beam widely used for heating in a DAC (Fig. 2). We first examined the EC of $H_2O$ ice VII in run #KO208 (Table 1). After compression to 26 GPa at room temperature, metallic electrodes were heated by lasers from both sides such that temperature of the $H_2O$ sample sandwiched in between was increased via thermal conduction. We simulated three-dimensional temperature distributions in a laser-heated sample by a finite element method and calculated the average temperature (= experimental temperature) of a cylindrical-shaped sample (see Supplementary Methods, Supplementary Figs. 1 and 2). In run #KO208, $H_2O$ ice was heated up to 772 K, which is below the melting temperature, by increasing the laser output power stepwise (see Methods). The impedance spectra were collected at each step (Fig. 3a), showing changes in sample resistance and accordingly the EC that is calculated from the resistance and the sample geometry (see Methods).

Subsequently we performed five separate sets of experiments to measure the direct-current electrical resistance and determine the EC of dense $H_2O$ fluid at 11–45 GPa and 1,030–2,750 K above the reported melting curve of $H_2O$ ice VII[12,13] (Fig. 1, Table 2). In run #KO221, the sample was first compressed to 36 GPa and then melted at 1,530 K by applying a given laser power output for about one second while continuously measuring the direct-current electrical resistance of the sample (Fig. 3b). We melted this sample again and obtained its resistance at higher temperatures of 2,020 K and 2,430 K. The sample EC at high P-T was obtained from the resistance and the geometry of the molten sample area based on simulations using a finite element method for each experiment (see Methods). Other four runs were carried out with collecting the direct-current electrical resistance of $H_2O$ fluid in a similar manner at varied high P-T conditions.

**Electrical conductivity of dense $H_2O$**

The measured EC of $H_2O$ ice and fluid was plotted as a function of reciprocal temperature in Fig. 4. For ice VII, we found that the EC is strongly temperature-dependent and was enhanced by three orders of magnitude with increasing temperature from 424 K to 772 K at 26 GPa. The following Arrhenius equation is fitted to such high P-T data;

$$\sigma = \sigma_0 \exp\left(-\frac{\Delta H}{kT}\right) \qquad (2)$$

where $\sigma_0$, $k$ and $\Delta H$ are a pre-exponential factor, Boltzmann constant and activation enthalpy, respectively. The fitting yields $\Delta H = 0.54(1)$ eV, suggesting that the conduction mechanism is dominated by proton conduction in $H_2O$ ice[19].

The EC found in this study at 300 K is consistent with the previously-reported room-temperature value at equivalent pressure by Okada et al.[20]. In contrast, it is approximately two orders of magnitude smaller than that obtained by Sugimura et al.[21] (Fig. 4). Such difference might be explained by overestimation of EC by the latter[21], in which two electrodes were placed on a single diamond anvil, extending from the sample hole to the culet and pavilion of the anvil. Therefore, a large area of the electrodes might have been in contact with excess ice that was present outside the sample chamber and contributed to electrical conduction. We also recognized that the EC at 300 K found in the present experiment is higher than the extrapolation of high-temperature data collected above 424 K (Fig. 4), implying a change in the conduction mechanism between these two temperatures. It is known that the electric charge in $H_2O$ ice is carried via proton conduction at ambient condition, which involves two types of defects; the ionic defect (the generation of $OH_3^+$ and $OH^-$ ions) and the orientational defect (the localization of hydrogen atoms by the rotation of water molecules) (see Figs. 5 and 6 in Ref.[22]). Both defects are thermally activated. Since the proton conduction requires both of these two defects, the defect with a smaller migration rate limits the proton conductivity[23]. The ionic defect concentration in $H_2O$ ice is four orders of magnitude lower than that of the orientational defects, and hence the ionic defect controls the EC[22] at ambient condition. On the other hand, molecular dynamics calculations[20] demonstrated that the orientational defect bounds the EC above 10 GPa at 800 K. The present experiments suggest that the rate-limiting process for the proton conduction in $H_2O$ ice VII changes from the ionic defect migration at 300 K to the orientational defect migration, at least above 424 K at 26 GPa.

We found the EC of dense $H_2O$ fluid to be ~10–40 S/m, higher by ~1.5 order of magnitude than that of $H_2O$ ice VII at melting temperature (Fig. 4). Our data on molten $H_2O$ collected in the pressure range from 11 and 45 GPa do not show the clear pressure dependence. Fitting Eq. 2 to the $H_2O$ fluid data gives $\Delta H$ = 0.14(7) eV. Previous shock experiments[4] observed a drastic increase in the EC of dense $H_2O$ fluid with increasing pressure to 25 GPa, which was attributed to the molecular-to-ionic transition in $H_2O$ fluid. However, we did not observe such an EC jump, indicating that the molecular-to-ionic transition in $H_2O$ fluid completes below 11 GPa and ~1,030 K, which is consistent with previous Raman spectroscopy measurements[24] suggesting the ionization of $H_2O$ above 12 GPa and 840 K.

Previous shock-compression data[2,5] showed the EC of ionic $H_2O$ fluid higher by about two orders of magnitude than our static measurements (Fig. 4). Such discrepancy may be attributed to overestimates in earlier studies due to large uncertainty in the geometry of a sample and electrodes during shock compression (note that the small activation enthalpy described above agrees with these shock compression measurements[2,5]). These shock experiments assumed that electrical conduction occurs only at the shock front, but Hamann and Linton[4] pointed out a possibility that sample behind the shock front also contributes to electrical conduction, which could cause the overestimation of the EC. In addition, any kinetic effects can affect the ionic conductivity measurements in nano- to microsecond timescales under dynamic compression. The theoretical calculations performed by French *et al.*[25] provided the ionic conductivity of $H_2O$ fluid that is consistent with our EC data within error bars, while the other calculations[26] showed about one order of magnitude higher conductivity.

We did not find the evidence of strong electronic conduction in $H_2O$ fluid up to 45 GPa and 2,750 K (Fig. 1). The earlier reflectivity measurements in shock experiments[15,17] demonstrated that the band gap of liquid $H_2O$ is ~6.5 eV at 1 bar and is still as wide as ~2 eV at 150 GPa[15], indicating no contribution of electronic conduction at *P-T* conditions explored in this study. Our results are in accordance with the conventional view that an ionic-to-electronic transition in dense $H_2O$ fluid takes place above ~4,000 K under tens of gigapascals. Indeed, recent static measurements of the absorption coefficient in a DAC reported the occurrence of electronic conduction at >17 GPa and >4,300 K[12], which is consistent with theoretical calculations[27].

**Implications for Uranus and Neptune interiors**
When considering conventional adiabatic temperature profiles[8] in the interiors of Uranus and Neptune, the electric charge in $H_2O$ fluid is transferred dominantly by ionic conduction rather than electronic conduction (Fig. 1). The present experiments found that the EC of ionic $H_2O$ fluid is ~10–40 S/m, two orders of magnitude lower than the previously believed value[28] of ~1,000 S/m based on earlier shock compression studies[2,5] (Fig. 4). The low EC of ionic $H_2O$ fluid observed in this study might suggest that convection in the ice giants is vigorous, which explains the existence of their magnetic fields. The EC of ~10–40 S/m requires the convection velocity of $U > $ ~10 cm/s to power a dynamo action by satisfying $R_m > 40$ (see Eq. 1). However, the convection velocity is proportional to a cubic root of planetary heat flux[29], and such vigorous convection is not viable especially for Uranus since its surface heat flux is low. It is likely that the convecting fluid is not pure $H_2O$ but includes some $NH_3$ and $CH_4$, but they are less conductive than ionic $H_2O$ and would not enhance the bulk conductivity[6].

Alternatively our results call for hotter interiors of the ice giants, high enough to evoke the electronic conduction in $H_2O$ fluid, at least around 4,300 K[12]. Indeed, recent simulations have proposed hotter interiors along with a thermal boundary layer developed within the mantle[10,11] due to compositional gradients or at the boundary between a stably-stratified H/He-rich envelope and the underlying dense $H_2O$-rich mantle[9] (Fig. 1). Previous simulations[11] of the temperature profile in the Neptune's interior suggested that a large part of a convective liquid $H_2O$ layer exceeds 4,300 K, in which $H_2O$ is electronically conducting with much higher EC than that of the ionic liquid[12], accounting for the generation of the magnetic field. Also, the hotter Uranus model proposed by Nettelmann *et al.*[9] showed >7,000 K for the convective $H_2O$-rich layer underneath the stratified H/He-rich envelope. Such hot ice giants' interiors might stabilize plasmatic $H_2O$ rather than electronically conducting fluid[9]. The role of such plasmatic $H_2O$ in generating the magnetic field has not been examined. It is important to note that the *P-T* condition for ionic-to-plasma transition has yet been poorly constrained[9] and the physical properties of plasmatic $H_2O$ are not known. Another hotter Uranus model proposed by Vazan and Helled[10] introduced a compositional gradient in the $H_2O$-rich mantle; a deeper part is not convecting and serves as a thermal boundary layer[30]. In their calculations, the temperature of the convective $H_2O$ layer does not reach 4,300 K[11] but exceeds 4,000 K, which could be high enough to activate electronic conduction in $H_2O$ fluid and generate a magnetic field.

Can such hot interior models explain the characteristic, multipolar magnetic fields of the Uranus and the Neptune? Based on the phase diagram of $H_2O$[12], the occurrence of electronically conducting $H_2O$ fluid layer suggests that the $H_2O$-rich mantle in these ice giants is fully molten (Fig. 1). It has been demonstrated[31] that the unique multipolar magnetic fields of Uranus and Neptune are produced by convection in a shallow, thin shell of ~0.2 times the radius of the icy giants from the surface, which is underlain by a non-convective layer. Indeed, earlier hot interior models[10,11] have already found that electronically conducting fluid layer would be separated into a shallower, thin convecting part and an underlying non-convecting portion, which likely produces the unique multipolar magnetic fields. Our finding of the EC of ionic $H_2O$ a few orders of magnitude lower than conventionally-used values supports such hot interiors of the ice giants, which gives valuable constraints on their interior structures[11] and thermal histories[9].

**Methods**

**High-pressure and -temperature experiments.** High *P-T* conditions were generated with laser-heated DAC techniques using flat anvils with 300 or 450 μm culet size. A several μm thick $Al_2O_3$ or $ZrO_2$ layer was sputtered onto the culet of both anvils (Fig. 2) by an Al or Zr target–$O_2$ gas reactive sputtering method using a radiofrequency magnetron sputtering equipment KS-701MS-TE1 (K. SCIENCE Corp.). Subsequently we also sputtered 500 nm thick Ir (electrode/laser absorber) and 50 nm thick Au (chemical insulator between the Ir and the $H_2O$ sample) on the $Al_2O_3$ or $ZrO_2$ layer using Quick Coater SC-701HMC II (SANYU ELECTRON Corp.). We employed a composite (rhenium + magnesia cement) gasket[18], which was ~300 μm thick before compression and had a hole (140 μm in diameter) as a sample chamber. The deposited Ir (+ Au) electrode was connected to a platinum foil put on both sides of the gasket, a copper wire and a source meter (Fig. 2). Before measuring the electrical resistance of the sample, we confirmed that a short circuit in this assembly showed the resistance of 0.1 ohm level, which defined the lower bound for the measurable resistance range.

We loaded a drop of deionized water into a sample chamber. Next, in order to avoid the outflow of $H_2O$ during compression, we put a whole DAC into a freezer for 30 min to make the sample solid. The sample was then compressed to a target pressure immediately after the DAC was taken out of the freezer. The sample temperature was increased by heating the Ir electrodes from both sides of the diamond anvils with a couple of 100 W single-mode Yb fiber lasers (YLR-100, IPG Photonics) (Fig. 2). We employed beam shapers that produce flat energy distributions of the laser beams. One-dimensional temperature profile on the surface of the Ir electrode was measured by a spectro-radiometric method[32]. Relatively low temperatures of the $H_2O$ ice sample in run #KO208 were estimated by extrapolation of the laser output power–temperature relation. The laser spot size was estimated from a collected radiation spectrum. We have simulated the temperature distributions in the $H_2O$ ice/fluid samples by a finite element method using a commercial software package COMSOL (see Supplementary Figs. 1 and 2). Temperature heterogeneity occurs in the molten $H_2O$ fluid sample, which depends on a laser spot size and sample thickness[18] but is not sensitive to the thermal conductivity of $H_2O$. For each experiment, temperature distributions were calculated based on the measured laser spot size and sample thickness. A cylindrical region of the laser-heated $H_2O$ sample was found to be molten in our fluid measurements. The temperature uncertainty may be ±10% typical for laser-heating experiments in a DAC. Pressure was determined based on the Raman spectrum of a diamond anvil at room temperature. We added the contribution of thermal pressure using an empirical value of 5% pressure increase for every 1,000 K temperature increase[33]. Errors in pressure may also be ±10%.

**Electrical conductivity measurements.** For $H_2O$ ice VII, the impedance spectra were collected at 26 GPa with increasing laser power output stepwise, using Chemical Impedance Analyzer 3532-80 (Hioki Corp.) and the Four-Terminal Probe 9500 (Hioki Corp.). The typical duration to collect a single impedance datum for $H_2O$ ice was 30 s at each temperature. The impedance of the system $Z(f) = R(f) + iX(f)$, where $R(f)$ and $X(f)$ are real and imaginary terms, respectively, were plotted in a Cole-Cole plot (Fig. 3a). Each spectrum was fitted with a parallel circuit of resistance $R$ and constant phase element to estimate the sample resistance using the commercial software Zview (Scribner Associates Inc., USA);

$$Z = \frac{R}{1+RC(2\pi i f)^p} \quad (3)$$

where $R$, $C$, $f$, $i$ and $p$ are resistance, capacitance, frequency, the imaginary unit and a fitting parameter ($p$ = 1 indicates an ideal capacitor), respectively. The uncertainty in $R$, $u_R$, was estimated from the fitting error, typically a few % (Table 1).

For liquid $H_2O$, we conducted direct-current (DC) resistance measurements at high temperatures during laser heating. Heating duration was limited to about one second to avoid possible chemical reaction between Au and the $H_2O$ sample. The voltage between the $V$ probes in a two-terminal assembly (Fig. 2) was measured using SourceMeter Keithley2450 (Keithley Corp.) under a constant current of 1 mA with 0.012% uncertainty. The sample resistance was calculated from the applied current and the measured voltage via Ohm's law. The sample resistance was obtained by reversing the direction of the applied current flow back and forth to remove the effect of the Seebeck voltage potentially caused by a temperature heterogeneity in a sample; we calculated the difference of the two readings of the sample resistance divided by two under current reversal to estimate the sample resistance (see the next section). Such sample resistance data were collected ~1,000 times within about one second in each heating cycle. The sample resistance was calculated by averaging all the resistance data collected during heating. The $u_R$ was estimated from the standard deviation of the collected sample resistance (Fig. 3b), which was below 1%.

The electrical conductivity $\sigma$ was calculated by;

$$\sigma = \frac{L}{RS} \quad (4)$$

where $L$ and $S$ are sample thickness and surface area, respectively. In this study, we assumed that the electric current by an applied electric field flows only in a laser-heated region, i.e., a molten cylindrical region of the sample. $L$ was determined by measuring on the cross section of a recovered sample (thickness of the insulating gasket) and then converted to thickness at high pressure using the equation of state of $H_2O$ phase VII[34], assuming isotropic sample volume change. $u_L$ was estimated from the standard deviation of the thickness of the insulating gasket. Note that the effect of a change in sample thickness upon heating can change the estimate of EC by at most a few % only, which is much smaller than a change by >~100 % in sample resistance with increasing temperature. $S$ at room temperature was estimated from the area where the electrode and the sample were directly in contact with each other. We consider the $S$ at high temperatures corresponds to the laser spot size. Its typical uncertainty $u_S$ was about 20%. The overall uncertainty in EC, $u_{EC}$, was calculated as;

$$u_{EC}=\sqrt{u_L^2 + u_S^2 + u_R^2} \tag{5}$$

**Sample resistance determined with current reversal technique.** We applied a specified current of ±1 mA to a sample and measured the voltage. Under the current sweep ($I_0 = +1$ mA), based on Ohm's rule, the following equation holds;

$$I_0 \times R_{sample} + \Delta V_S = V_{meas+} \tag{6}$$

where $R_{sample}$ is sample resistance, $\Delta V_S$ is the Seebeck voltage, and $V_{meas+}$ is measured voltage, respectively.

Upon current reversal, the direction of the current changes but the temperature gradient field stays constant to the first approximation, and so the Ohm's rule can be written as;

$$-I_0 \times R_{sample} + \Delta V_S = V_{meas-} \tag{7}$$

where $V_{meas-}$ is measured voltage upon current reversal. By removing $\Delta V_S$ from Eqs. 6 and 7,

$$2 \times I_0 \times R_{sample} = V_{meas+} - V_{meas-} \tag{8}$$

$$R_{sample} = (V_{meas+} - V_{meas-}) / (2 \times I_0) \tag{9}$$

Therefore, the effect of Seebeck voltage can be canceled out by calculating the difference between the two readings of the measured resistance ($V_{meas+}/I_0$, $V_{meas-}/I_0$) divided by two under the current reversal.

**Simulation of temperature distributions in a heated sample.** Temperature distributions in the $H_2O$ ice VII and $H_2O$ fluid samples during laser heating were simulated by a finite element method using the software package COMSOL Multiphysics (COMSOL Inc.) (Supplementary Figs. 1 and 2). We employed the measured thickness of the sample, being sandwiched between 3 μm thick alumina or zirconia layers. We have not included the Ir+Au electrodes in the calculations because their thickness was only a few microns or less in total. A homogeneous circular area $S$ with a diameter of the laser spot size was considered on both sample surfaces (Tables 1 and 2), and the applied heat was conducted to the sample and surrounding alumina or zirconia layers. The back surface of the diamond anvils in contact with air and WC seats was set at room temperature. The interface between the gasket and air was considered to be a thermal insulating boundary. The thermal conductivity of $H_2O$ was assumed to be 20 W/m/K based on room temperature literature data[35]. There are no high-temperature thermal conductivity data but we found the thermal conductivity of $H_2O$ negligibly affects temperature distributions. Indeed, the use of an order of magnitude lower thermal conductivity of $H_2O$ did not remarkably change the simulation results (Supplementary Fig. 2c). The thermal conductivity of the thermal insulator was set at 10 W/m/K, which is a typical value for corundum under high *P-T* conditions[36]. Tetrahedral meshes with a resolution of 0.1 µm were applied.

We performed the simulation for each run and estimated a three-dimensional average temperature of a heated sample (= experimental temperature). The difference between the measured temperature on Ir electrodes and the estimated sample average temperature is provided in Supplementary Table 1.

**Acknowledgements**
We thank Hisayoshi Shimizu for valuable discussion. This work was supported by the JSPS grant 21H04968 to K.H.

**Author contributions**
K.O. performed the experiments with help by Y.O. and K.H. K.O. and Y.O. conceptualized the research. K.H., Y.O. and K.O. wrote the manuscript.

**Competing interests**
The authors declare no competing interests.

**Data and materials availability**
All data are provided in the main text or the supplementary materials. Additional data supporting the results of this study are available from the corresponding author upon reasonable request.

# Figures and tables

Table 1 | **Experimental results on H$_2$O ice VII**

| Laser output power (W) | Tempertaure (K) | Resistance $R$ (Ohm) | $R$ error (%) | Capacitance $C$ (pF) | Fitting parameter $p$ | Conductivity $\log_{10}[\sigma\ (S/m)]$ |
|---|---|---|---|---|---|---|
| 0  | 300     | $1.93 \times 10^8$ | 4 | 12.6(9)  | 0.862(6)  | -4.96(+6/-7)  |
| 12 | 424(42) | $1.32 \times 10^8$ | 3 | 8.0(3)   | 0.918(3)  | -3.44(+8/-10) |
| 14 | 454(45) | $6.00 \times 10^8$ | 3 | 8.9(6)   | 0.908(6)  | -3.01(+8/-10) |
| 16 | 476(48) | $2.86 \times 10^7$ | 2 | 8.2(4)   | 0.914(4)  | -2.77(+8/-10) |
| 18 | 499(50) | $1.42 \times 10^7$ | 1 | 7.2(4)   | 0.928(4)  | -2.47(+8/-10) |
| 20 | 528(53) | $7.04 \times 10^6$ | 2 | 8.4(7)   | 0.915(6)  | -2.16(+8/-10) |
| 22 | 550(55) | $4.02 \times 10^6$ | 2 | 7.5(10)  | 0.925(10) | -1.92(+8/-10) |
| 24 | 573(57) | $2.48 \times 10^6$ | 2 | 4.8(7)   | 0.956(11) | -1.71(+8/-10) |
| 26 | 602(60) | $1.67 \times 10^6$ | 2 | 4.4(6)   | 0.965(10) | -1.54(+8/-10) |
| 28 | 624(62) | $1.02 \times 10^6$ | 2 | 7.5(10)  | 0.928(10) | -1.33(+8/-10) |
| 30 | 647(65) | $6.97 \times 10^5$ | 3 | 8.0(21)  | 0.919(18) | -1.16(+8/-10) |
| 32 | 676(68) | $4.69 \times 10^5$ | 2 | 6.7(13)  | 0.938(13) | -0.99(+8/-10) |
| 34 | 698(70) | $4.04 \times 10^5$ | 2 | 4.9(10)  | 0.959(13) | -0.92(+8/-10) |
| 36 | 721(72) | $3.69 \times 10^5$ | 1 | 5.4(9)   | 0.950(12) | -0.88(+8/-10) |
| 38 | 750(75) | $2.50 \times 10^5$ | 2 | 5.1(9)   | 0.955(11) | -0.71(+8/-10) |
| 40 | 772(77) | $1.89 \times 10^5$ | 2 | 10.8(38) | 0.906(23) | -0.59(+8/-10) |

Sample thickness $L$ was estimated to be 20(2) μm. Sample area $S$ at ambient temperature was measured to be 9,500(950) μm$^2$ from the area of electrode in contact with the sample, and 414(76) μm$^2$ at high temperature from the laser spot size.

**Table 2 | Experimental results on H₂O fluid at high *P-T***

| Run # | Temperature (K) | Pressure (GPa) | Area $S$ (μm²) | Length $L$ (μm) | Resistance $R$ (Ohm) | Conductivity $\sigma$ (S/m) |
|---|---|---|---|---|---|---|
| KO217 | 1,030(110) | 11(1) | 260(26) | 7.3(1) | 3,410(70) | 8.2(8) |
|  | 1,250(130) | 12(1) | 260(26) | 7.3(1) | 2,550(30) | 11.0(11) |
|  | 1,470(150) | 12(1) | 260(26) | 7.3(1) | 1,480(40) | 18.2(20) |
| KO218 | 1,610(160) | 19(2) | 501(50) | 13.0(3) | 2,330(130) | 11.2(13) |
| KO220 | 1,545(150) | 24(2) | 394(39) | 13.8(5) | 1,350(30) | 26.0(26) |
|  | 1,497(150) | 24(2) | 325(33) | 13.8(5) | 1,090(50) | 39.0(43) |
|  | 1,445(140) | 36(3) | 320(32) | 13.4(5) | 2,030(30) | 20.7(21) |
| KO221 | 1,530(150) | 38(4) | 222(22) | 12.8(2) | 1,940(30) | 29.7(30) |
|  | 2,020(200) | 42(4) | 443(44) | 12.8(2) | 1,590(50) | 18.1(19) |
|  | 2,430(240) | 45(4) | 401(40) | 12.7(2) | 1,980(160) | 16.1(21) |
| KO222 | 2,750(280) | 25(2) | 384(38) | 24.7(3) | 1,760(40) | 36.6(37) |

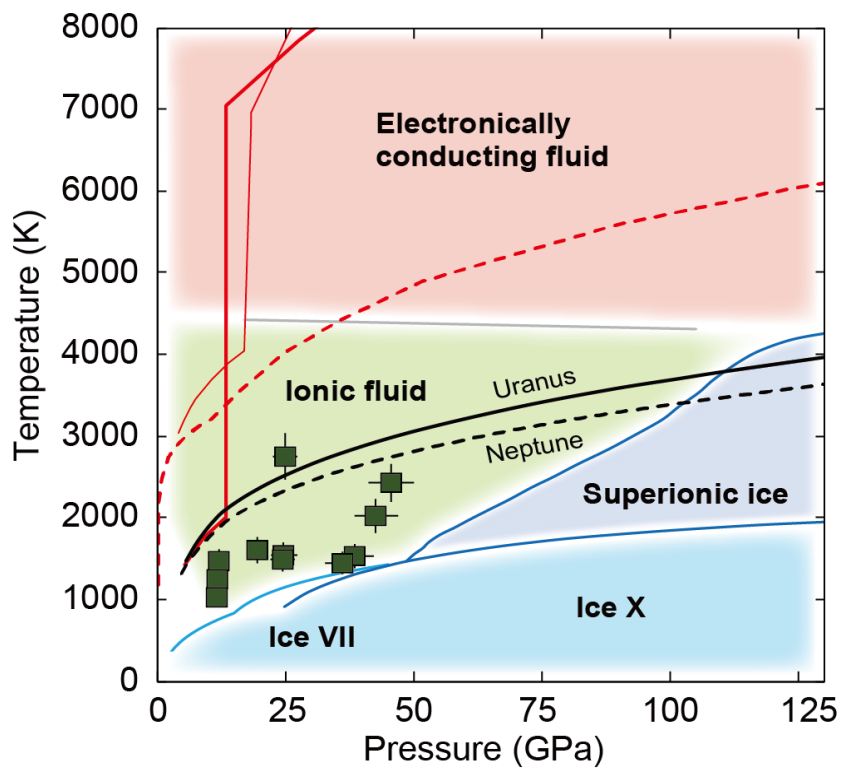

**Fig. 1 | Phase diagram of $H_2O$ and the present experimental conditions.** Green symbols indicate high *P-T* conditions of the EC measurements in this study. Light and dark blue lines show phase boundaries previously reported by experimental[37] and theoretical studies[8], respectively. A grey line is the boundary between ionic and electronically conducting fluids[12]. The conventional adiabatic temperature profiles of the interiors of Uranus and Neptune are given by solid and dashed black lines, respectively[8]. The hotter thermal structures are shown by red curves; bold[9] and thin[10] solid lines for Uranus and a dashed line[11] for Neptune.

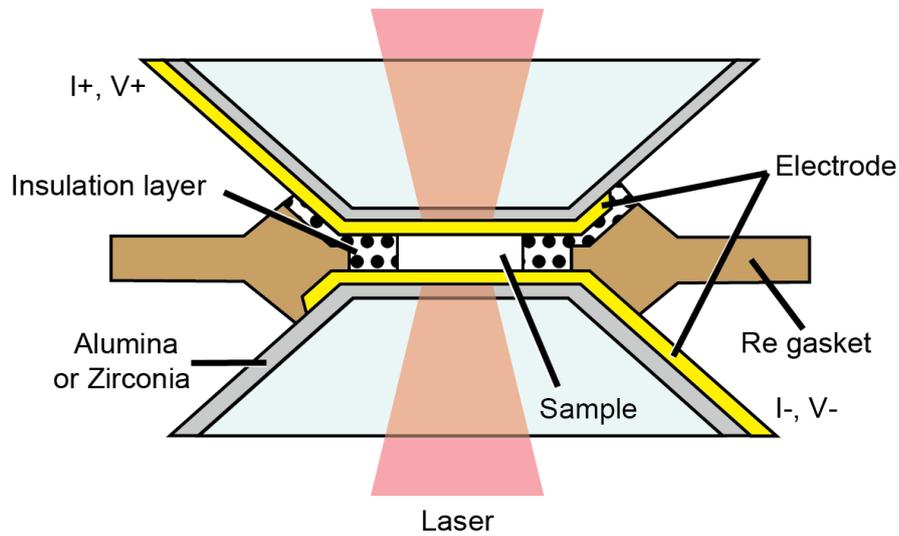

**Fig. 2 | Configuration in a laser-heated DAC designed for electrical conductivity measurements of $H_2O$.** Thin thermal insulation layers of alumina or zirconia were deposited on diamond anvils using a reactive sputtering technique (grey). Iridium and gold were then sputtered on top of the alumina or zirconia layers as a laser absorber and an electrode for electrical resistance measurement. Gasket was composed of (outer) rhenium and (inner) electrical insulation layer of magnesia cement. $H_2O$ was loaded into a sample chamber as water and then frozen into ice at -10 °C before compression to minimize deformation of a sample chamber.

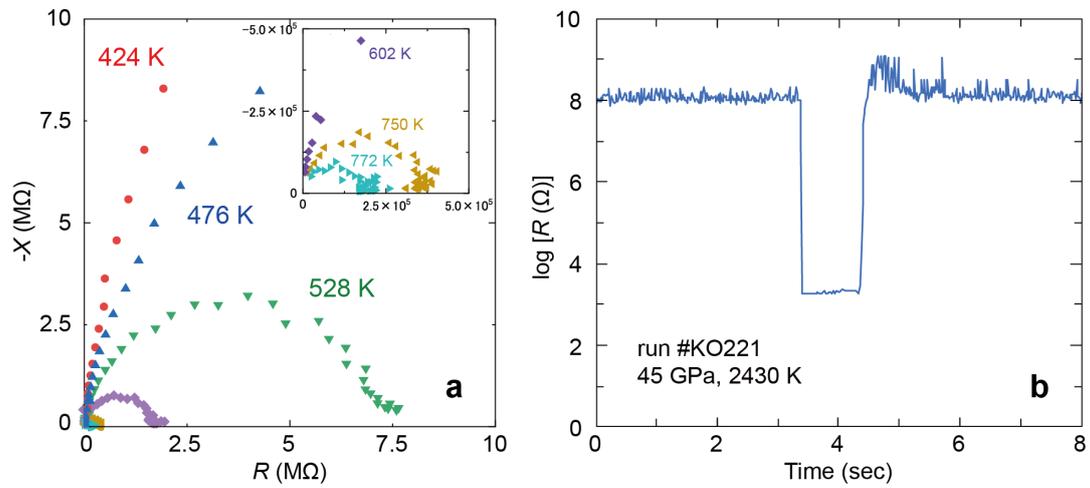

**Fig. 3 | Electrical resistance data for H₂O ice VII and fluid collected at high *P-T*. a,** The impedance spectra of H₂O ice from 424 K to 772 K at 26 GPa in run #KO208. **b,** Time variations in direct-current electrical resistance of molten H₂O before/during/after laser heating to 2,430 K at 45 GPa in run #KO221. The resistance suddenly decreased at the onset of laser heating and returned back to original upon quenching temperature.

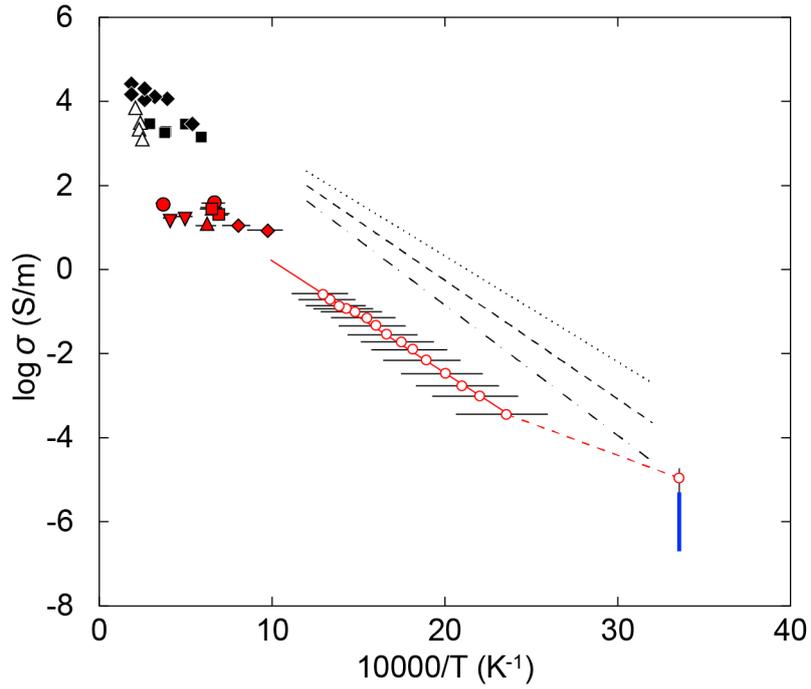

**Fig. 4 | The EC of dense $H_2O$ at high *P-T*.** Red symbols show the present determinations for ice VII (open circles) and fluid (closed symbols), respectively. A blue bar is the EC of ice VII at room temperature determined by Okada *et al.*[20]. The dotted, broken and dot-broken lines indicate earlier EC data of $H_2O$ ice at 20, 40 and 60 GPa, respectively[21]. Solid squares and diamonds are shock compression data for ionic $H_2O$ ionic fluid[2,5], and open triangles are static compression data for electronically conducting $H_2O$ fluid[12].

# Supplementary Material

# Hot interiors of ice giant planets inferred from electrical conductivity of dense $H_2O$ fluid

Kenta Oka[1], Yoshiyuki Okuda[1]* & Kei Hirose[1,2]

[1]Department of Earth and Planetary Science, The University of Tokyo, Tokyo, Japan
[2]Earth-Life Science Institute, Tokyo Institute of Technology, Tokyo, Japan

**Supplementary Figures**

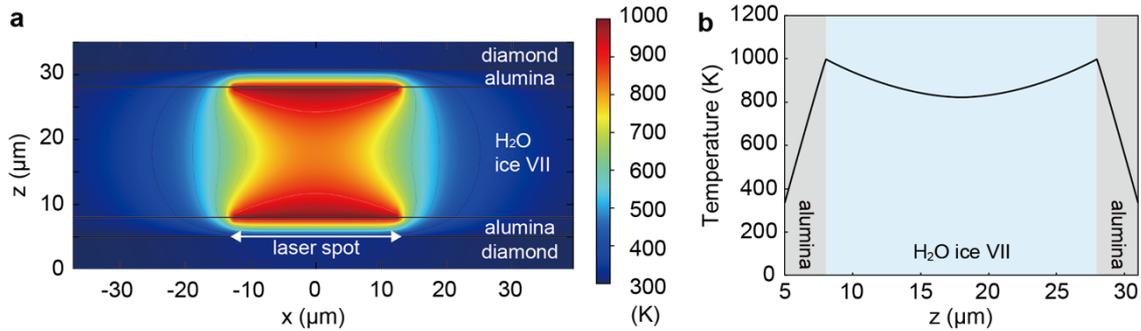

**Supplementary Fig. 1 | Simulated temperature distributions in solid $H_2O$ ice heated to an average temperature of 772 K in run #KO208. a,** 2D cross-sectional result. **b,** 1D temperature profile along the z-axis at x = 0 in **a**. Thin curves in **a** indicate isotherms with 100 K intervals.

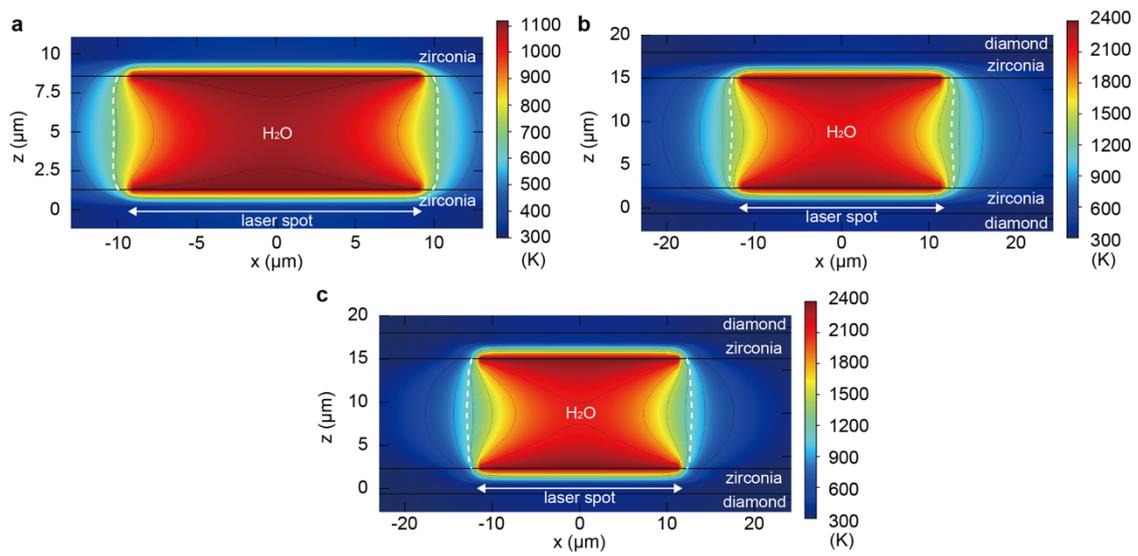

**Supplementary Fig. 2 | Simulated 2D cross-sectional temperature distributions of $H_2O$ fluid. a,** run #KO217 at 11 GPa heated to an average temperature of 1,030 K. **b, c,** run #KO221 at 41 GPa and an average temperature of 2,020 K. The thermal conductivity of $H_2O$ is considered to be 20 W/m/K in **a, b** and 2 W/m/K in **c**. Thin curves indicate isotherms with 200 K intervals in **a** and 300 K intervals in **b** and **c**. White broken curves show the isotherm of the melting temperature of $H_2O$ at each pressure[37].

# Supplementary Table

**Supplementary Table 1 | Measured temperature of the Ir electrode surface and the simulated average sample temperature**

| Run # | Ir surface (K) | Sample (K) |
|---|---|---|
| KO208 | 300 | 300 |
|  | 470 | 424(42) |
|  | 510 | 454(45) |
|  | 540 | 476(48) |
|  | 570 | 499(50) |
|  | 610 | 528(53) |
|  | 640 | 550(55) |
|  | 670 | 573(57) |
|  | 710 | 602(60) |
|  | 740 | 624(62) |
|  | 770 | 647(65) |
|  | 810 | 676(68) |
|  | 840 | 698(70) |
|  | 870 | 721(72) |
|  | 910 | 750(75) |
|  | 940 | 772(77) |
| KO217 | 1,120 | 1,030(110) |
|  | 1,370 | 1,250(130) |
|  | 1,610 | 1,470(150) |
| KO218 | 1,870 | 1,610(160) |
| KO220 | 1,860 | 1,545(150) |
|  | 1,840 | 1,497(150) |
|  | 1,760 | 1,445(140) |
| KO221 | 1,940 | 1,530(150) |
|  | 2,390 | 2,020(200) |
|  | 2,920 | 2,430(240) |
| KO222 | 4,060 | 2,750(280) |